\documentclass[11pt]{article}

\usepackage{amsmath}
\usepackage{epsfig}
\usepackage{graphics}

\newcommand{\N}{N\raise.7ex\hbox{\underline{$\circ $}}$\;$}

\begin{document}

\title{
 V.M. Red'kov\footnote{redkov@dragon.bas-net.by}\\
Ricci  Coefficients in Covariant   Dirac Equation,\\ Symmetry
Aspects and Newman-Penrose Approach}

\maketitle

\begin{quotation}

The paper  investigates how  the Ricci rotation coefficients act
in the Dirac equation in presence of external gravitational fields
described in terms of Riemannian space-time geometry. It is shown
that only 8 different combinations of the Ricci coeffici\-ents
$\gamma_{abc}(x)$ are involved  in the Dirac equation. They are
combined in two 4-vectors $B_{a}(x)$ and $C_{a}(x)$ under local
Lorentz group which has  status of the gauge symmetry group. In
all orthogonal coordinates  one  of these vectors,
"pseudovector"\hspace{2mm} $C_{a}(x)$,  vanishes identically. The
gauge transformation laws of the two vectors are found explicitly.
Connection of these   $B_{a}(x)$ and $A_{a}(x)$ with the  known
Newman-Penrose coefficients is established. General study of
gauge symmetry aspects in Newman-Penrose formalism is performed.
 Decomposition of the Ricci object,  "tensor" \hspace{2mm}
 $\gamma_{abc}(x)$,  into two "spinors"\hspace{2mm} $\gamma(x)$  and $\bar{\gamma}(x)$
 is done. At this Ricci rotation coefficients are divided into two groups:
12 complex functions  $\gamma(x) = ( \gamma^{\alpha}_{\;\;\beta\dot{\rho}\sigma})$
and 12 conjugated to them $\bar{\gamma}(x) =
( \gamma_{\dot{\alpha}}^{\;\;\dot{\beta}\rho\dot{\sigma}})$.
Components of spinor $\bar{\gamma}(x)$ coincide with
12  spin coefficients by Newman-Penrose
$
\kappa \; , \; \pi \; , \; \epsilon \; , \; \rho \; , \; \lambda \; ,
\; \alpha \; , \;  \sigma \; , \; \beta \; , \; \tau \; , \; \nu \; , \; \gamma \; .
$
For listing  these it is  used a special letter-notation
 $L_{i}\; , \; N_{i} \; , \; M_{i} \; , \; \bar{M}_{i}\; $.
The  formulas for gauge transformations of spin
coefficients under local Lorentz group are derived.
There are given two solutions  to the gauge problem: one in the compact form of transformation
laws for spinors $\gamma(x)$  and $\bar{\gamma}(x)$, and another as
detailed elaboration
of the latter in terms of 12 spin coefficients.

\end{quotation}

\section{Dirac equation and Ricci  coefficients}

 The known  Dirac equation on the background of a
curved space-time involves  the Ricci rotation coefficients [1],
the  non-linear objects of a curved space-time geometry, and it
has the form [2-4] ($A_{\alpha}(x)$ stands for an external
electromagnetic field, $c=1, \hbar =1$)

\begin{eqnarray}
 \left\{   \gamma ^{\alpha } \;
\left[\; i  \left( {\partial \over x^{\alpha }} + \Gamma _{\alpha } \right) -
e  A_{\alpha}\;  \right]  -
 m     \right\} \Psi  = 0  \; , \;\;\;
\nonumber
\\
\gamma^{\alpha}(x) = \gamma^{b} e_{(b)}^{\alpha}(x),\;\;  \Gamma
_{\alpha }(x) = {1 \over 2} \sigma ^{ab}  e^{\beta }_{(a)}
 e_{(b)\beta ; \alpha } \; ,\;\;
 \sigma^{ab} = {1\over 4} (\gamma^{a} \gamma^{b} - \gamma^{b} \gamma^{a}) \; ,
\label{1.1}
\end{eqnarray}

\noindent or
\begin{eqnarray}
 \left \{   \gamma^{c}  [  i
( e_{(c)}^{\alpha}   \partial_{\alpha}  +
{1\over 2} \sigma^{ab} \gamma_{abc}    ) - e  A_{c}    ]
 -   m    \right  \} \Psi  = 0
\label{1.2}
\end{eqnarray}

\noindent where  $\gamma_{abc}(x)$ are the Ricci rotation symbols [1]
\begin{eqnarray}
\gamma_{bac} (x)= -\;  \gamma_{abc} (x) =
-\;  e_{(b)\beta ; \alpha }  \;  e_{(a)}^{\beta} e_{(c)}^{\alpha}
\label{1.3}
\end{eqnarray}

\noindent and $A_{a} (x) = e_{(a)}^{\alpha}(x)  A_{\alpha}(x)  $  designates tetrad (vierbein) components
of the  electromagnetic field 4-vector.
 Now, with the  use of the known formula for product of three Dirac matrices [5]
\begin{eqnarray}
\gamma^{c} \gamma^{a} \gamma^{b} = \gamma^{c} g^{ab} - \gamma^{a}
g^{cb} + \gamma^{b} g^{ca} + i \gamma^{5} \epsilon ^{cabk}
\gamma_{k}  \; , \qquad
\gamma^{5} = - i \gamma^{0} \gamma^{1}
\gamma^{2} \gamma^{3}\;  , \qquad \epsilon^{0123} = +1 \;
\nonumber
\end{eqnarray}

\noindent we can easily produce
\begin{eqnarray}
\gamma^{c} \sigma^{ab} =
{1 \over 2}\; (g^{ca} \gamma^{b} - g^{cb} \gamma^{a}  + i \gamma^{5}
\; \epsilon^{cabk} \gamma_{k} ) \; .
\label{1.4}
\end{eqnarray}

\noindent  Taking in mind eq. (\ref{1.4}),  eq. (\ref{1.2}) can be  transformed into
\begin{eqnarray}
\left \{   \; \gamma^{k}\;  [\;  i \; ( e_{(k)}^{\alpha}  \;
\partial_{\alpha}  + {1\over 2} \; e_{(k);\alpha}^{\alpha}    - {i
\over 4} \;\gamma^{5} \epsilon ^{abc}_{\;\;\;\;\;\;k}
\gamma_{abc}  )
 -  e  A_{c} ] - m  \; \right \} \Psi  = 0 \; .
\label{1.5}
\end{eqnarray}

With the notation
\begin{eqnarray}
B_{k}(x) =   {1 \over 2}  \; e_{(k)\; ;\alpha }^{\alpha}(x)  \; ,
\;\;\; \qquad C_{k}(x) = {1 \over 4}
\;\epsilon^{abc}_{\;\;\;\;\;\;k}  \;  \gamma_{abc}(x) \;
\label{1.6}
\end{eqnarray}

\noindent
the Dirac equation (\ref{1.1}) will get the form
\begin{eqnarray}
\left \{  \gamma^{k} \; [ i
( e_{(k)}^{\alpha} \partial_{\alpha}  +   B_{k}
 -  i \gamma^{5} C_{k}  ) -  e  A_{a}  ]
 -  m \right  \}  \Psi  = 0  .
\label{1.7}
\end{eqnarray}

This form of the Dirac equation is remarkable  in some  aspects.
The first one is that the vector field $C_{k}(x)$ involved in eq. (\ref{1.7})  vanishes identically in all
orthogonal coordinates and their accompanying tetrads.
Therefore,  the  Dirac equation  will take on the simpler form
\begin{eqnarray}
\left \{  \gamma^{k} \; [\; i
( e_{(k)}^{\alpha} \partial_{\alpha}  +   B_{k}
) -  e  A_{a}   \; ] -   m  \right \}  \Psi   = 0    \; .
\label{1.8}
\end{eqnarray}

Let  us prove this property of the vector $C_{k}(x)$.
By definition,
$
\gamma_{abc} = -\gamma_{bac}  \; ;
$
for the following is is useful  to introduce  a quantity antisymmetric with respect to $bc$:
\begin{eqnarray}
\lambda_{abc}(x) = \gamma_{abc} (x)- \gamma_{acb}(x) \;  .
\nonumber
\end{eqnarray}

\noindent For  $\lambda_{abc} (x) $
there exists representation in terms of ordinary derivatives:
\begin{eqnarray}
\lambda_{abc} = \gamma_{abc} - \gamma_{acb} =
( e_{(a)\alpha ; \beta } - e_{(a)\beta ; \alpha } )
e_{(c)}^{\alpha} e_{(b)}^{\beta} \;\;\;
\nonumber
\\
= ( \partial_{\beta} e_{(a)\alpha} - \Gamma^{\rho}_{\alpha \beta} e_{(a)\rho} -
 \partial_{\alpha} e_{(a)\beta} + \Gamma^{\rho}_{\beta \alpha} e_{(a)\rho} )
e_{(c)}^{\alpha} e_{(b)}^{\beta}
\nonumber
\\
=  ( \partial_{\beta} e_{(a)\alpha}  -  \partial_{\alpha} e_{(a)\beta} )
e_{(c)}^{\alpha} \;e_{(b)}^{\beta} \; . \qquad \qquad
\nonumber
\end{eqnarray}

\noindent Also the  identity
\begin{eqnarray}
\gamma_{abc} = {1 \over 2} \;(\lambda_{abc} + \lambda_{bca} - \lambda_{cab} )
= {1 \over 2}\; ( \gamma_{abc} - \gamma_{acb} + \gamma_{bca} - \gamma_{bac} -
\gamma_{cab} + \gamma_{cba} )
\label{1.9}
\end{eqnarray}

\noindent holds. With the use of eqs. (\ref{1.9}), for components of
$C_{k}(x)$  (\ref{1.6}) it follows
\begin{eqnarray}
C_{0}(x) = \;\; \epsilon^{abc}_{\;\;\;\;\;\;0} \; \gamma_{abc}(x) =
(\lambda_{123} + \lambda_{231} +\lambda_{312} ) \; ,
\nonumber
\\[2mm]
C_{1}(x) =  \epsilon^{abc}_{\;\;\;\;\;\;1}  \; \gamma_{abc}(x) =
 - (\lambda_{203} + \lambda_{302} +\lambda_{023} ) \; ,
\nonumber
\\[2mm]
C_{2}(x) = \;\; \epsilon^{abc}_{\;\;\;\;\;\;2} \; \gamma_{abc}(x) =
(\lambda_{301} + \lambda_{013} +\lambda_{130} ) \; ,
\nonumber
\\[2mm]
C_{3}(x) =  \epsilon^{abc}_{\;\;\;\;\;\;3}  \; \gamma_{abc}(x) =
-(\lambda_{012} + \lambda_{120} +\lambda_{201} ) \; . \label{1.10}
\end{eqnarray}

Let us  consider  these relations (\ref{1.10}) in a space-time with
a diagonal metric tensor:
\begin{eqnarray}
dS^{2} =
h_{0}^{2} (x)\; (dx^{0})^{2} \;- \;h_{i}^{2}(x) \; (dx^{i})^{2} \; ,
\nonumber
\end{eqnarray}

\noindent and its accompanying tetrad
\begin{eqnarray}
e_{(a) \alpha}(x) = \left ( \begin{array}{cccc}
h_{0} & 0 & 0 & 0 \\
0  &  h_{1} & 0 & 0 \\
0  &  0 & h_{2} & 0 \\
0  &  0  & 0 &   h_{3}
\end{array} \right )     .
\label{1.11}
\end{eqnarray}

Taking into account  (\ref{1.11}) and
 (\ref{1.9}), for eq. (\ref{1.10}) we  can easily obtain the form
\begin{eqnarray}
C_{0}(x) =\;\;
[ \partial_{2} e_{(1)3} - \partial_{3} e_{(1)2} ]
e^{2}_{(2)}e^{3}_{(3)} +
[ \partial_{3} e_{(2)1}
-\;\partial_{1} e_{(2)3}  ]
e^{3}_{(3)}e^{1}_{(1)} +
[  \partial_{1} e_{(3)2} - \partial_{2} e_{(3)1}  ]
e^{1}_{(1)}e^{2}_{(2)} \; ,
\nonumber
\\[2mm]
C_{1}(x) =
 - [  \partial_{3} e_{(2)0} - \partial_{0} e_{(2)3}  ]
e^{3}_{(3)}e^{0}_{(0)} -
[ \partial_{0}  e_{(3)2}
- \;\partial_{2}  e_{(3)0}  ]
e^{0}_{(0)}e^{2}_{(2)} - [  \partial_{2}  e_{(0)3} - \partial_{3}  e_{(0)2}  ]
e^{2}_{(2)}e^{3}_{(3)} \; ,
\nonumber
\\[2mm]
C_{2}(x) =\;\;
  [  \partial_{0}  e_{(3)1} - \partial_{1}  e_{(3)0}  ]
e^{0}_{(0)}e^{1}_{(1)} +
[  \partial_{1} e_{(0)3}
-\; \partial_{3} e_{(0)1}  ]
e^{1}_{(1)}e^{3}_{(3)} +
[  \partial_{3} e_{(1)0} - \partial_{0}  e_{(1)3}  ]
e^{3}_{(3)}e^{1}_{(1)} \; ,
\nonumber
\\[2mm]
C_{3}(x) =
 - [  \partial_{1}  e_{(0)2} - \partial_{2}  e_{(0)1}  ]
e^{1}_{(1)}e^{2}_{(2)} -
[  \partial_{2} e_{(1)0}
-\; \partial_{0}  e_{(1)2}  ]
e^{2}_{(2)}e^{0}_{(0)} -
[  \partial_{0} e_{(2)1} - \partial_{1}  e_{(2)0}  ]
e^{0}_{(0)}e^{1}_{(1)} \; .
\nonumber
\end{eqnarray}

\noindent
It should be noted that  these  relations involve only
non-diagonal elements  of  the tetrad matrix $e_{(a)\beta}(x)$, therefore
all quantities $C_{k}(x)$ vanish identically.
So, always in any space-time models characterized by (\ref{1.11})
the above vector combination of the Ricci coefficient is zero:
\begin{eqnarray}
C_{k}(x) = {1 \over 4} \; \epsilon^{abc}_{\;\;\;\;\;\;k}  \;  \gamma_{abc}(x)
\equiv 0 \; .
\label{1.12}
\end{eqnarray}

We are to give attention to separation
of these 8 relevant constituents  from $\gamma_{abc}$.
To this end, as the first step, one decomposes $\gamma _{abc}(x)$ into the sum
\begin{eqnarray}
\gamma _{abc} = [ \gamma _{abc}  -
 A  \epsilon_{abc}^{\;\;\;\;\;\;n} \; C_{n}(x)  ] + A  \epsilon_{abc}^{\;\;\;\;\;\;n}  C_{n}(x)
\nonumber
\\[2mm]
\equiv
  \; \Delta_{[ab]c}(x) +
A \; \epsilon_{abc}^{\;\;\;\;\;n} \; C_{n}(x)  \; \;  \qquad  \qquad
\label{1.13a}
\end{eqnarray}

\noindent  where  $A$  is to be chosen  later. Indeed, let us  require
\begin{eqnarray}
\epsilon^{abc}_{\;\;\;\;\;m} \Delta_{[ab]c}(x) =0 \; 
\qquad \mbox{or}
 \qquad
4 C_{m}(x) -  A \; ( \epsilon^{abc}_{\;\;\;\;\;m}
\epsilon_{abc}^{\;\;\;\;\;n} )\; C_{n}(x)\; =0. \qquad \nonumber
\end{eqnarray}

\noindent
From this, with relation
$
\epsilon^{abc}_{\;\;\;\;\;m}
\epsilon_{abc}^{\;\;\;\;\;n} =- 6 \delta_{m}^{n}
$, it follows
\begin{eqnarray}
  (4  + 6 A)\; C_{m} \; =0 
  \qquad
\mbox{or} 
\qquad A = - {2 \over 3} \; .
\nonumber
\end{eqnarray}

\noindent
With the  use of the following notation for 3-rank tensor dual to a vector $C_{n}$:
\begin{eqnarray}
C_{[abc]}(x) = -{2 \over 3} \; \epsilon_{abc}^{\;\;\;\;\;\;n} \; C_{n}(x)
\label{1.13b}
\end{eqnarray}

\noindent  the expansion  (\ref{1.13a})  looks as
\begin{eqnarray}
\gamma _{abc}(x) =    \Delta_{[ab]c}(x) + C_{[abc]}(x)
 \; .
\label{1.13c}
\end{eqnarray}

The latter provides us with the decomposition of the $\gamma_{abc}(x)$
into  the sum of $\Delta_{[ab]c}(x)$, orthogonal to the  Levi-Civita symbol
\begin{eqnarray}
\epsilon^{abc}_{\;\;\;\;\;\;m} \Delta_{[ab]c}(x) = 0 \; ,
\label{1.14a}
\end{eqnarray}

\noindent  and
the $C_{[abc]}(x)$ non-orthogonal to the Levi-Civita symbol
\begin{eqnarray}
\epsilon^{abc}_{\;\;\;\;\;\;m}\; C_{[abc]}(x) =
 \epsilon^{abc}_{\;\;\;\;\;\;m}  \left [
 -{2 \over 3} \; \epsilon_{abc}^{\;\;\;\;\;\;n} \; C_{n}(x) \right ]
 =
 + 4 \; C_{m}(x) = \epsilon^{abc}_{\;\;\;\;\;\;m} \;  \gamma_{abc}(x)  \; . \qquad
\label{1.14b}
\end{eqnarray}

Further, taking in mind the known formula
$$
\epsilon^{klm}_{\;\;\;\;\;\;n} \epsilon_{abc}^{\;\;\;\;\;\;n} =
(-1) \left ( \begin{array}{llll}
\delta^{k}_{a}  & \delta^{k}_{b}  &  \delta^{k}_{c} \\
\delta^{l}_{a}  & \delta^{l}_{b}  &  \delta^{l}_{c} \\
\delta^{m}_{a}  & \delta^{m}_{b}  &  \delta^{m}_{c}
\end{array} \right )
$$

\noindent one  produces the following  form of  $C_{[abc]}(x)$  in terms of the
Ricci
coefficients:
\begin{eqnarray}
C_{[abc]}(x) = -{2 \over 3} \; \epsilon_{abc}^{\;\;\;\;\;\;n} \;
C_{n}(x) = -{2 \over 3} \; \epsilon_{abc}^{\;\;\;\;\;\;n} \;
 {1 \over 4} \;  \epsilon^{klm}_{\;\;\;\;\;\;\;n}  \;  \gamma_{klm}(x)
={1 \over 3} \; (\gamma_{abc} + \gamma_{bca} + \gamma_{cab} )\; .
\nonumber
\end{eqnarray}

\noindent In addition, for $\Delta_{abc}$ one  gets to
\begin{eqnarray}
\Delta_{[ab]c} (x) = \gamma_{abc} - C_{[abc]} = {2 \over 3} \;
\gamma_{abc} + {1 \over 3} \;\left  ( \gamma_{acb}(x) -
\gamma_{bca}(x) \right ) \; . \label{1.15}
\end{eqnarray}

Take notice that one cannot obtain
from $C_{[abc]}(x)$, by means of simplification over  any pair
of indices,  a non-zero vector. In other words, this tensor is irreducible.
But such a  trick is possible with the  $\Delta_{[ab]c}(x)$:
\begin{eqnarray}
\Delta_{[ab]c}(x) = [  \Delta_{[ab]c}(x) -
\alpha   (  g_{ac} B_{b}(x) - g_{bc} B_{a}(x)  )  ]
\nonumber
\\[2mm]
+ \alpha   ( g_{ac} B_{b}(x) - g_{bc} B_{a}(x) )   \equiv
 E_{[ab]c}(x) +  B_{[ab]c}(x),
 \nonumber
 \end{eqnarray}

\noindent where
\begin{eqnarray}
B_{b}(x) = \gamma_{kb}^{\;\;\;\;k} (x) = - \gamma_{b\;\;\;\;k}^{\;\;k} \; ,
\nonumber
\\[2mm]
B_{[ab]c}(x)   = \alpha \; ( g_{ac} B_{b}(x) - g_{bc} B_{a}(x))    \; ,
\nonumber
\\[2mm]
E_{[ab]c} = \Delta_{[ab]c}(x) - B_{[ab]c}(x) \; .
\label{1.16b}
\end{eqnarray}

\noindent The choice  $\alpha = +1 / 3$ insures the properties
\begin{eqnarray}
B_{[kb]}^{\;\;\;\;\;k}(x) = B_{b}(x) \; , \;\;
B_{[bk]}^{\;\;\;\;\;k}(x) = - B_{b}(x) \; ,
\nonumber
\\[2mm]
E_{[kb]}^{\;\;\;\;\;k}(x) = 0 \; , \qquad
E_{[bk]}^{\;\;\;\;\;k}(x) = 0  \; . \qquad
\label{1.16c}
\end{eqnarray}

\noindent Besides, the quantity $E_{[ab]c}(x)$ is orthogonal to the  Levi-Civita symbol:
\begin{eqnarray}
\epsilon^{abc}_{\;\;\;\;\;\;m} E_{[ab]c}(x) = 0 \; .
\label{1.17}
\end{eqnarray}

So, we get to the result we need: the  Ricci object can be composed as the sum
\begin{eqnarray}
\gamma_{abc}(x) = C_{[abc]}(x) + B_{[ab]c}(x) + E_{[ab]c}(x)
\label{1.18}
\end{eqnarray}

\noindent where  the tensors  $C_{[abc]}(x)$ and  $B_{[ab]c}(x)$
are the combinations which are relevant as we concern the Dirac equation in any
curved space-time model. Besides, in any orthogonal coordinate system
the tensor $C_{[abc]}(x)$  vanishes identically.

\section{Gauge properties of  $B_{a}$ and $C_{a}$}

\hspace{5mm}
Now we are going to consider how the above two vector fields $B_{a}(x)$ and $C_{a}(x)$
behave  with respect to any local  tetrad Lorentz transformation  [1]. For more generality
we will take account of  proper as well as improper Lorentz matrices
\begin{eqnarray}
e_{(a)}^{'\beta}(x) = L_{a}^{\;\; b} (x)  \; e_{(b)}^{\beta} (x)
\; , \qquad L_{a}^{\;\; b} (x) = L_{a}^{\;\; b} (k(x),k^{*}(x)) \;
. \;\;\;\; \label {2.1}
\end{eqnarray}

\noindent
Starting from definition of  $B_{a}(x)$,  one can readily produce
\begin{eqnarray}
B'_{a}(x) = \nabla_{\beta}  \left [\;  e_{(a)}^{'\beta}(x) \;
\right ] = \nabla_{\beta}  \left [ \; L_{a}^{\;\;b}(x) \;
e_{(b)}^{\beta} \; \right ] =L_{a}^{\;\;b} (x) \; (\;
\nabla_{\beta} \;  e_{(b)}^{\beta} \;) \; +\; {\partial
L_{a}^{\;\;b}  \over \partial x^{\beta}}  \;e_{(b)}^{\beta} \qquad
\nonumber
\end{eqnarray}

\noindent from where it follows the transformation law  for $B_{a}(x)$ we need:
\begin{eqnarray}
B'_{a}(x) =     L_{a}^{\;\;b} (x) \; B_{b}(x) \; + \;
{\partial L_{a}^{\;\;b}(x)   \over \partial x^{\beta} }\; \;
 e_{(b)}^{\beta} \; .
\label {2.2}
\end{eqnarray}

Analogously let us analyze the case of  $C_{a}(x)$. Now, it will be convenient to go
from the  known formulas for gauge transformation of the Ricci coefficients [1]
\begin{eqnarray}
\gamma'_{abc}(x) = L_{a}^{\;\;k}(x)  L_{b}^{\;\;l} (x)
L_{c}^{\;\;n}(x)  \; \gamma_{kln}(x) + \; L_{a}^{\;\;k} (x)\;
g_{kl} \; { \partial L_{b}^{\;\;l}(x) \over \partial x^{\mu}} \;
L_{c}^{\;\; n}(x)  \; e_{(n)}^{\mu}(x) \; . \label {2.3}
\end{eqnarray}

 Instead of $L_{a}^{\;\;k}(x) g_{kl}$ we will write $L_{al}(x) $ and so on;
with this notation the orthogonality condition for Lorentz  matrices will take the form
$L_{ab} = L^{-1}_{ba}$.
Multiplying eq. (\ref{2.3}) by ${1 \over 4} \epsilon^{abc}_{\;\;\;\;\;\;d}$,
we get to
\begin{eqnarray}
C'_{d}(x) = {1 \over 4} \; \epsilon^{abc}_{\;\;\;\;\;\;d} \;
L_{a}^{\;\;k} (x) L_{b}^{\;\;l}(x)  L_{c}^{\;\;n} (x) \;
\gamma_{kln}(x)
\nonumber
\\
 +{1 \over 4} \; \epsilon^{abc}_{\;\;\;\;\;\;d}\;
L_{al}(x) \; { \partial L_{b}^{\;\;l}(x) \over \partial x^{\mu}}
\; L_{c}^{\;\; n}(x) \;  e_{(n)}^{\mu}(x) \; .
\label {2.4}
\end{eqnarray}

Now, taking into account the known formula
$$
\epsilon^{abcd} \; L_{a}^{\;\;k}  L_{b}^{\;\;l}  L_{c}^{\;\;n}  L_{d}^{\;\;m} =
+ \;\mbox{det} \;[L_{s}^{\;\;t} ] \; \; \epsilon^{klnm}\; ,
$$

\noindent  we will have
$$
\epsilon^{abc}_{\;\;\;\;\;\;d} \; L_{a}^{\;\;k}  L_{b}^{\;\;l}  L_{c}^{\;\;n}
= + \;\mbox{det}\;  [L_{s}^{\;\;t}]\;  \epsilon^{klnm}\; L_{dm} \;   ,
$$

\noindent with the  use of which in eq. (\ref{2.4}) we get to the
required gauge law ($\mbox{det}\; (L_{s}^{\;\;t}) = \mbox{det} \; L $):
\begin{eqnarray}
C'_{d}(x) = \mbox{det} L(x) \;\;  L_{d}^{\;\; m}(x)  \; C_{m}(x)
\;
\nonumber
\\
 + \; {1 \over 4}\; \epsilon^{abc}_{\;\;\;\;\;\;d} \left [
L_{al}(x) { \partial (L^{-1})^{l}_{\;\;b}(x) \over
\partial x^{\mu} } \right ]  \;  L_{c}^{\;\;n}(x)\;
e_{(n)}^{\mu}(x) \; .
\label {2.5}
\end{eqnarray}

\section {Connection with the  Newman-Penrose  spin formalism}

Now we are going to dwell upon  the structure of the Dirac
equation (\ref{1.7}) in  a detailed component-based
form\footnote{In the widely used  method of spin coefficients by
Newman-Penrose [4] just a such approach is exploited; we consider
the above Dirac equation (\ref{1.7})  in that  spin-coefficient
language and then compare it with the form  based on the  use of
the vectors $B_{a}(x)$ and $C_{a}(x)$ (\ref{1.6}).}.
 As a  first
step let us write  down the Dirac equation in  the 2-spinor
(splitted) form (the conventional notation for spinors based on
dotted and undotted indices is used)
\begin{eqnarray}
\sigma^{a} \; [ \; i \; (e_{(a)}^{\beta} \partial_{\beta} + B_{a} +
i C_{a}  ) -e  A_{a} \; ] \; \xi = m \; \eta \; ,
\nonumber
\\
\bar{\sigma}^{a} \; [\; i\;  (e_{(a)}^{\beta} \partial_{\beta} + B_{a} -
i C_{a}  ) - e  A_{a} \; ] \; \eta = m \; \xi  \; .
\label {3.1}
\end{eqnarray}

\noindent From this, allowing for the explicit form of the Pauli
two-by-two matrices, and introducing  a special letter designation
according to
\begin{eqnarray}
B_{(0)} + B_{(3)} = \hat{B}_{0} \; ,  \qquad  B_{(0)} - B_{(3)} =
\hat{B}_{1} \; , \nonumber
\\
B_{(1)} -iB_{(2)} =\hat{B}_{2} \; , \qquad
B_{(1)} + i B_{(2)} = \hat{B}_{3} \; ,
\nonumber
\\
C_{(0)} + C_{(3)} = \hat{C}_{0} \; , \qquad
C_{(0)} - C_{(3)} = \hat{C}_{1} \; ,
\nonumber
\end{eqnarray}
\begin{eqnarray}
C_{(1)} -iC_{(2)} =\hat{C}_{2} \; , \qquad  C_{(1)} + i C_{(2)} =
\hat{C}_{3} \; , \nonumber
\\
A_{(0)} + A_{(3)} = \hat{A}_{0} \; , \qquad  A_{(0)} - A_{(3)} =
\hat{A}_{1} \; , \nonumber
\\
A_{(1)} -iA_{(2)} = \hat{A}_{2} \; ,\qquad  A_{(1)} + i A_{(2)} =
\hat{A}_{3} \; , \nonumber
\\
e_{(0)}^{\beta} + e_{(3)}^{\beta} = \hat{e}_{0}^{\beta} \; ,
\qquad  e_{(0)}^{\beta} - e_{(3)}^{\beta} = \hat{e}_{1}^{\beta} \;
, \nonumber
\\
e_{(1)}^{\beta} -i e_{(2)}^{\beta} =\hat{e}_{2}^{\beta} \; ,
\qquad  e_{(1)}^{\beta} + i e_{(2)}^{\beta} = \hat{e}_{3}^{\beta}
\; ,
\nonumber
\end{eqnarray}

\noindent for eqs.  (\ref{3.1}) we  get the form
\begin{eqnarray}
\sigma^{a} \; [ \; i\;   (e_{(a)}^{\beta} \partial_{\beta} + B_{a} +
i C_{a}  ) - e  A_{a} \; ] \; \xi
\nonumber
\\
= \left ( \begin{array}{l}
i ( \hat{e}_{0}^{\beta} \partial_{\beta}  +    \hat{B}_{0}
+ i   \hat{C}_{0} ) - e  \hat{A}_{0}  \\
 i  ( \hat{e}_{3}^{\beta} \partial_{\beta}  +   \hat{B}_{3}
+ i   \hat{C}_{3} ) - e  \hat{A}_{3}
\end{array} \right. \qquad
\left. \begin{array}{l} i( \hat{e}_{2}^{\beta} \partial_{\beta}  +
\hat{B}_{2}
+ i   \hat{C}_{2} ) - e  \hat{A}_{2}   \\
 i  ( \hat{e}_{1}^{\beta} \partial_{\beta}  +    \hat{B}_{1}
+ i   \hat{C}_{1} ) - e  \hat{A}_{1}
\end{array} \right ) \xi =m \eta \; ,
\nonumber
\\[5mm]
\bar{\sigma}^{a} \; [\; i\;  (e_{(a)}^{\beta} \partial_{\beta} + B_{a} +
i C_{a}  ) - e  A_{a} \; ] \; \eta
\nonumber
\\
= \left ( \begin{array}{r}
i ( \hat{e}_{1}^{\beta} \partial_{\beta}  +    \hat{B}_{1}
- i   \hat{C}_{1} ) - e  \hat{A}_{1}  \\
- i  ( \hat{e}_{2}^{\beta} \partial_{\beta}  +   \hat{B}_{3} - i
\hat{C}_{3} ) + e  \hat{A}_{3}   \end{array} \right. \qquad
\left.
\begin{array}{r} - i( \hat{e}_{3}^{\beta} \partial_{\beta}  +
\hat{B}_{2}
- i   \hat{C}_{2} ) + e  \hat{A}_{2}   \\[0.2cm]
 i  ( \hat{e}_{0}^{\beta} \partial_{\beta}  +    \hat{B}_{0}
- i   \hat{C}_{0} ) - e  \hat{A}_{0}
\end{array} \right )
\eta = m\;  \xi \; .
\nonumber
\\
\label{3.3}
\end{eqnarray}

\noindent The forms obtained reflect explicitly that only 8  of
the 24 Ricci coefficients are involved in the  Dirac equation: $
\hat{B}_{0},\hat{B}_{1},\hat{B}_{2},\hat{B}_{3}$
        and $ \hat{C}_{0},\hat{C}_{1},\hat{C}_{2},\hat{C}_{3}$.

Now we will go into the  notation accepted in the Newman-Penrose approach [4].
First, for describing the connection  $B_{\alpha}(x)$  in spinor basis one is to
introduce the following  designation
\begin{eqnarray}
B_{\alpha} (x) =
\left (   \begin{array}{cc}
\Sigma_{\alpha} & 0 \\ 0  & \bar{\Sigma}_{\alpha}
\end{array} \right )
=  {1 \over 8} \left ( \begin{array}{cc}
\bar{\sigma}^{\beta} \sigma_{\beta ;\alpha} -
\bar{\sigma}_{\beta ;\alpha} \sigma^{\beta}  &  0 \\
0   &   \sigma^{\beta}  \bar{\sigma}_{\beta ;\alpha} -
\sigma_{\beta ;\alpha} \bar{\sigma}^{\beta}
\end{array} \right ),
\label{3.4}
\end{eqnarray}

\noindent where
\begin{eqnarray}
\sigma_{\beta}(x) = \sigma^{a} e_{(a) \beta} (x) \; , \qquad
\bar{\sigma}_{\beta} (x)  = \bar{\sigma}^{a} e_{(a)\beta}(x) \;   ,
\nonumber
\\[0.3cm]
\sigma_{\beta ; \alpha}(x) = \sigma^{a} e_{(a)\beta ; \alpha}  \;, \qquad
\bar{\sigma}_{\beta ; \alpha} = \bar{\sigma}^{a} e_{(a) \beta ; \alpha}  \; .
\nonumber
\end{eqnarray}

\noindent
Also,  one is to employ  the letter notation for elements of the  matrices
  $\sigma^{\beta}(x)$  and $\bar{\sigma}^{\beta}(x)$:
\begin{eqnarray}
\sigma^{\beta}(x) =
\left ( \begin{array}{cc}
 e_{(0)}^{\beta}  +  e_{(3)}^{\beta}   &  e_{(1)}^{\beta}  -i e_{(2)}^{\beta}  \\
 e_{(1)}^{\beta}  + i e_{(2)}^{\beta}  &  e_{(0)}^{\beta}  -  e_{(3)}^{\beta}
\end{array} \right )
=
\sqrt{2}
\left ( \begin{array}{cc}
l^{\beta}(x)  &  m^{\beta} (x) \\
\bar{m}^{\beta}(x) & n^{\beta}(x)
\end{array} \right ) \; ,
\nonumber
\\
\bar{\sigma}^{\beta}(x) =
\left ( \begin{array}{cc}
e_{(0)}^{\beta}  -  e_{(3)}^{\beta}    & - e_{(1)}^{\beta}  +i e_{(2)}^{\beta}  \\
- e_{(1)}^{\beta}  - i e_{(2)}^{\beta}  &  e_{(0)}^{\beta}   +  e_{(3)}^{\beta}
\end{array} \right )
=
\sqrt{2}
\left ( \begin{array}{cc}
n^{\beta}(x)  &  -m^{\beta}(x) \\
-\bar{m}^{\beta} (x) & l^{\beta}(x)
\end{array} \right ),
\label{3.5}
\end{eqnarray}

\noindent  and further
$$
\bar{\sigma}^{\beta} \sigma_{\beta ;\alpha}
=
2 \left ( \begin{array}{rr}
n^{\beta} l_{\beta ; \alpha} - m^{\beta} \bar{m}_{\beta ; \alpha} &
n^{\beta} m_{\beta ; \alpha} - m^{\beta} n_{\beta ; \alpha} \\
- \bar{m}^{\beta} l_{\beta ; \alpha} + l^{\beta} \bar{m}_{\beta ; \alpha} &
- \bar{m}^{\beta} m_{\beta ; \alpha} + l^{\beta} n_{\beta ; \alpha}
\end{array} \right )  \; ,
$$
$$
\bar{\sigma}_{\beta ; \alpha} \sigma^{\beta}
=
 2 \left ( \begin{array}{rr}
n_{\beta ; \alpha} l^{\beta}  - m_{\beta ; \alpha} \bar{m}^{\beta} &
n_{\beta ; \alpha} m^{\beta}  - m_{\beta ; \alpha} n^{\beta} \\
-\bar{m}_{\beta ; \alpha} l^{\beta}  + l_{\beta ; \alpha} \bar{m}^{\beta} &
-\bar{m}_{\beta ; \alpha} m^{\beta}  + l_{\beta ; \alpha} n^{\beta}
\end{array} \right ) \; .
$$

\noindent
So, the  expression for connection $\Sigma_{\alpha}(x)$ is
\begin{eqnarray}
2\Sigma_{\alpha}
\label{3.6}
= \left ( \begin{array}{cc}
-  l^{\beta} n_{\beta ; \alpha } -
m^{\beta} \bar{m}_{\beta ; \alpha}
& 2\; n^{\beta} m_{\beta ; \alpha} \\
2\; l^{\beta} \bar{m}_{\beta ; \alpha}
&
l^{\beta} n_{\beta ; \alpha } +
m^{\beta} \bar{m}_{\beta ; \alpha}
\end{array} \right ) .
\end{eqnarray}

\noindent  In getting  (\ref{3.6}) one  must allow for that
arbitrary generally covariant scalar products
$$
l^{\beta}(x), n^{\beta}(x), m^{\beta}(x),
\bar{m}^{\beta}(x)$$
are generally covariant invariants,
therefore the  identities of the form
\begin{eqnarray}
n^{\beta} l_{\beta} = inv \Longrightarrow
n^{\beta}_{\;\;;\alpha} l_{\beta} + n^{\beta} l_{\beta ; \alpha} = 0 \; ,
\nonumber
\\
n^{\beta} m_{\beta} = inv \Longrightarrow
n^{\beta}_{\;\;;\alpha} m_{\beta} + n^{\beta} m_{\beta ; \alpha} = 0 \; ,
\nonumber
\\
n^{\beta} \bar{m}_{\beta} = inv \Longrightarrow
n^{\beta}_{\;\;;\alpha} \bar{m}_{\beta} + n^{\beta} \bar{m}_{\beta ; \alpha} = 0\;\;
\nonumber
\end{eqnarray}

\noindent and so on hold.
In the same manner we  find the expression for
$\bar{\Sigma}_{\alpha}(x)$-connection:
$$
\sigma^{\beta} \bar{\sigma}_{\beta ;\alpha} =
2
\left ( \begin{array}{rr}
l^{\beta}  &  m^{\beta} \\
\bar{m}^{\beta} & n^{\beta}
\end{array} \right )
\left ( \begin{array}{rr}
n_{\beta ;\alpha}  &  -m_{\beta ; \alpha} \\
-\bar{m}_{\beta ; \alpha} & l_{\beta ; \alpha}
\end{array} \right )
$$
$$
= 2
\left ( \begin{array}{rr}
l^{\beta} n_{\beta ; \alpha} - m^{\beta} \bar{m}_{\beta ; \alpha} &
-l^{\beta} m_{\beta ; \alpha} + m^{\beta} l_{\beta ; \alpha} \\
\bar{m}^{\beta} n_{\beta ; \alpha} - n^{\beta} \bar{m}_{\beta ; \alpha} &
- \bar{m}^{\beta} m_{\beta ; \alpha} + n^{\beta} l_{\beta ; \alpha}
\end{array} \right ) ,
$$
$$
\sigma_{\beta ; \alpha} \bar{\sigma}^{\beta}   =
2
\left ( \begin{array}{rr}
l_{\beta ; \alpha}  &  m_{\beta ; \alpha} \\
\bar{m}_{\beta ; \alpha} & n_{\beta ; \alpha}
\end{array} \right )
\left ( \begin{array}{cc}
n^{\beta}  &  -m^{\beta} \\
-\bar{m}^{\beta} & l^{\beta}
\end{array} \right )
$$
$$
=2 \left ( \begin{array}{rr}
l_{\beta ; \alpha} n^{\beta}  - m_{\beta ; \alpha} \bar{m}^{\beta} &
-l_{\beta ; \alpha} m^{\beta}  + m_{\beta ; \alpha} l^{\beta} \\
\bar{m}_{\beta ; \alpha} n^{\beta}  - n_{\beta ; \alpha} \bar{m}^{\beta} &
-\bar{m}_{\beta ; \alpha} m^{\beta}  + n_{\beta ; \alpha} l^{\beta}
\end{array} \right )  ,
$$

\noindent
and finally
\begin{eqnarray}
2\; \bar{\Sigma}_{\alpha}
\label{3.7}
= \left ( \begin{array}{cc}
  l^{\beta} n_{\beta ; \alpha } +
\bar{m}^{\beta} m_{\beta ; \alpha} &
-2 \; l^{\beta} m_{\beta ; \alpha} \\
-2\; n^{\beta} \bar{m}_{\beta ; \alpha}  &
- l^{\beta} n_{\beta ; \alpha } -
\bar{m}^{\beta} m_{\beta ; \alpha}
\end{array} \right ).
\end{eqnarray}

To write down  the Dirac equation
\begin{eqnarray}
i\; \sigma ^{\alpha }(x) \; ( \partial _{\alpha }
 \; + \; \Sigma _{\alpha }(x) ) \; \xi (x) = \; m \; \eta (x)  \;   ,
\nonumber
\\
i\; \bar{\sigma }^{\alpha }(x) \; ( \partial _{\alpha } \; + \;
\bar{\Sigma }_{\alpha }(x) ) \; \eta (x) = \; m \; \xi (x) \; ,
\nonumber
\end{eqnarray}

\noindent  in the Newman-Penrose approach we are to develop
expressions for involved differential operators in corresponding notation.
To this  end we   obtain
$$
i\; \sigma ^{\alpha }(x) \; ( \partial _{\alpha }
 \; + \; \Sigma _{\alpha }(x) )
= i \sqrt{2}  \; \left [ \;
\left ( \begin{array}{cc}
l^{\alpha} \partial_{\alpha} & m^{\alpha} \partial_{\alpha} \\
\bar{m}^{\alpha} \partial_{\alpha} & n^{\alpha} \partial_{\alpha}
\end{array} \right )  \right.
$$
$$
+
\left ( \begin{array}{l}
 - {1\over 2} (l^{\beta}n_{\beta ; \alpha} + m^{\beta} \bar{m}_{\beta;\alpha})
l^{\alpha} + m^{\alpha} l^{\beta}  \bar{m}_{\beta;\alpha}  \\[0.2cm]
- {1\over 2} (l^{\beta}n_{\beta ; \alpha} +
m^{\beta} \bar{m}_{\beta;\alpha})\bar{m}^{\alpha} + n^{\alpha} l^{\beta}
\bar{m}_{\beta;\alpha}
\end{array} \right.
\left. \left. \begin{array}{l}
  + {1\over 2} (l^{\beta}n_{\beta ; \alpha} +
m^{\beta} \bar{m}_{\beta;\alpha})m^{\alpha} + l^{\alpha}n^{\beta} m_{\beta;\alpha}
\\[0.2cm]
+ {1\over 2} (l^{\beta}n_{\beta ; \alpha} +
m^{\beta} \bar{m}_{\beta;\alpha})n^{\alpha} + \bar{m}^{\alpha} n^{\beta} m_{\beta;\alpha}
\end{array}  \right )   \right  ]  .
$$

\noindent Let us  introduce conventions:
\begin{eqnarray}
l^{\alpha} \partial_{\alpha} = \nabla_{l}\; , \qquad
n^{\alpha} \partial_{\alpha} = \nabla_{n}\; , \;
\nonumber
\\
m^{\alpha} \partial_{\alpha} =  \nabla_{m} \; , \qquad
\bar{m}^{\alpha} \partial_{\alpha} =  \nabla_{\bar{m}} \; ;
\nonumber
\end{eqnarray}

\noindent besides it will be convenient to omit pairs of mute indices:
\begin{eqnarray}
\bar{m}_{\beta;\alpha}  l^{\beta}  m^{\alpha}  =
\bar{m}  l  m  = - l \bar{m} m \; , \;\;
\nonumber
\\
 - {1\over 2} ( n_{\beta ; \alpha} l^{\beta} +
 \bar{m}_{\beta;\alpha}m^{\beta} )
l^{\alpha} =
 - {1\over 2} ( n l + \bar{m} m) l
\nonumber
\end{eqnarray}

\noindent  and so on. Thus, the above equation  will look as
\begin{eqnarray}
i \sigma ^{\alpha }(x)  ( \partial _{\alpha }
  +  \Sigma _{\alpha }(x) )
= \left ( \begin{array}{l}
\nabla_{l}  - {1\over 2} (n l +  \bar{m} m) l + \bar{m} l m  \\
\nabla_{\bar{m}} - {1\over 2} ( n l +  \bar{m} m) \bar{m} + \bar{m} l n
\end{array}  \right.
\left. \begin{array}{r}
 \nabla_{m} + {1\over 2} (n l +  \bar{m} m) m +  m n l  \\
\nabla_{n} + {1\over 2} ( n l +  \bar{m} m ) n +  m n   \bar{m}
\end{array}  \right )
.
\label{3.8}
\end{eqnarray}

\noindent In the same manner we find
\begin{eqnarray}
i\bar{\sigma} ^{\alpha }(x)  ( \partial _{\alpha }
  +  \bar{\Sigma} _{\alpha }(x) )
=
\left ( \begin{array}{l}
\nabla_{n} + {1\over 2} (n l + m \bar{m}) n  + \bar{m} n m   \\
- \nabla_{\bar{m}} - {1\over 2} (n l + m \bar{m} ) \bar{m} - \bar{m}nl
\end{array}
 \begin{array}{l}
-\nabla_{m}  + {1\over 2} (n l + m  \bar{m}) m - m l n\\
\nabla_{l} + m l \bar{m} - {1\over 2} (n l + m \bar{m}) l
\end{array}  \right ) .
\label{3.9}
\end{eqnarray}

Let us define 12 complex combinations of the Ricci coefficients:
\begin{eqnarray}
\left. \begin{array}{rr}
(l \bar{m}) l  =  L_{1}  &
( l \bar{m} ) m  =  M_{1} \\
(- n m ) l = L_{2} &
(- n m ) m =  M_{2}    \\
{1 \over 2} ( l  n  +  m  \bar{m} ) l  =  L_{3} &
{1 \over 2} ( l  n  +  m  \bar{m} )  m = M_{3}  \\[2mm]
(l \bar{m}) \bar{m}  =  \bar{M}_{1}  &
( l \bar{m} ) n  =  N_{1}  \\
(- n m  )  \bar{m} =  \bar{M}_{2}     &
( - n  m )   n =  N_{2}         \\
{1 \over 2} (l n  + m \bar{m} ) \bar{m}\;  =  \bar{M}_{3} &
{1 \over 2} (l n + m  \bar{m})  n = N_{3}
\end{array} \right.
\label{3.10}
\end{eqnarray}

\noindent These $( L_{i}, \; N_{i}, \;M_{i}, \;\bar{M}_{i} )$
are related straightforwardly with the so-called Newman-Penrose coefficients [4]
$$
( k , \;\pi  , \; \epsilon  ; \;
\rho  , \; \lambda  , \; \alpha  ; \; \sigma  , \; \mu , \; \beta  ; \;
\tau , \; \nu , \; \gamma  )
$$

\noindent  according to
$$
\left. \begin{array}{ccc}
L_{1} \;  = \;   \; k^{*}    \qquad     &
L_{2} \;  = \;   \; \pi ^{*} \qquad     &
L_{3} \;  = \;   \; \epsilon ^{*}      \\
N_{1} \;  = \;   \; \tau^{*} \qquad     &
N_{2} \;  = \;   \; \nu ^{*} \qquad     &
N_{3} \;  = \;   \; \gamma ^{*}        \\
M_{1} \;  = \;   \; \rho ^{*} \qquad    &
M_{2} \;  = \;   \; \lambda ^{*} \qquad &
M_{3} \;  = \;   \; \alpha ^{*}        \\
\bar{M}_{1} \; = \;   \; \sigma ^{*} \qquad  &
\bar{M}_{2} \; = \;   \; \mu ^{*} \qquad     &
\bar{M}_{3} \; = \;   \; \beta ^{*}\;.
\end{array} \right.
$$

\noindent
In  this notation, the  above differential operators read as
\begin{eqnarray}
i \sigma^{\alpha}(x) (\partial_{\alpha} + \Sigma_{\alpha}(x)) = i \sqrt{2}
 \left ( \begin{array}{ll}
(\nabla_{l} + \;\;L_{3} - M_{1})  & (\nabla_{m} + L_{2} - M_{3}) \\
(\nabla_{\bar{m}} + \bar{M}_{3} - N_{1})  & (\nabla_{n} + \bar{M}_{2} - N_{3})
\end{array} \right )   ,
\label{3.11}
\end{eqnarray}
\begin{eqnarray}
i \bar{\sigma}^{\alpha}(x) (\partial_{\alpha} + \bar{\Sigma}_{\alpha}(x)) = i \sqrt{2}
\left ( \begin{array}{rr}
(\nabla_{n} + \bar{M}^{*}_{2} - N^{*}_{3}) &
 -(\nabla_{m} + \bar{M}^{*}_{3} - N^{*}_{1}) \\
-(\nabla_{\bar{m}} + L^{*}_{2} - M^{*}_{3}) &
(\nabla_{l} + \;\;L^{*}_{3} - M^{*}_{1})
\end{array} \right )  .
\label{3.12}
\end{eqnarray}

It should be  noted that in eqs. (\ref{3.11}) and (\ref{3.12}) only the following 8 spin
 coefficients and their conjugates
\begin{eqnarray}
\left. \begin{array}{llllllll}
L_{2}, & L_{3}, & \bar{M}_{2}, &  \bar{M}_{3} ,  &   N_{1},  & N_{3} , &   M_{1}, & M_{3}, \\
L^{*}_{2}, & L^{*}_{3}, &  \bar{M}^{*}_{2}, &  \bar{M}^{*}_{3}, &
 N^{*}_{1},  & N^{*}_{3},  &   M^{*}_{1}, &  M^{*}_{3}  ,
\end{array} \right.
\nonumber
\end{eqnarray}

\noindent  are involved, and  what is more, these quantities enter the
equations only in
combinations
\begin{eqnarray}
L_{3} - M_{1}   ,  \quad  L_{2} - M_{3}  ,\quad
\bar{M}_{3} - N_{1}  ,  \nonumber
\\
\bar{M}_{2} - N_{3} ,\quad
(L_{3} - M_{1})^{*}   , \quad   (L_{2} - M_{3})^{*}  ,\nonumber
\\
(\bar{M}_{3} - N_{1})^{*}  , \quad    (\bar{M}_{2} - N_{3})^{*}\;  .
\nonumber
\end{eqnarray}

\noindent
In other words, this means that only 8 real-valued combinations of the Ricci
object
enter the Dirac equation.

It remains to establish how these spin-coefficients-based parameters
refer to the  above vectors $C_{a}(x)$ and $B_{a}(x)$ (\ref{1.6}).
By straightforward comparison  we derive
\begin{eqnarray}
\hat{B}_{0} + i \hat{C}_{0} = \sqrt{2}\;  (L_{3} -M_{1}) \; ,
\nonumber
\\[0.2cm]
\hat{B}_{0} - i \hat{C}_{0} = \sqrt{2}  \; (L^{*}_{3} - M^{*}_{1}) \; ,
\nonumber
\\[0.2cm]
\hat{B}_{1} + i \hat{C}_{1} = \sqrt{2} \; (\bar{M}_{2} - N_{3}) \; ,
\nonumber
\\[0.2cm]
\hat{B}_{1} - i \hat{C}_{1} = \sqrt{2}  \; (\bar{M}^{*}_{2} -N^{*}_{3}) \; ,
\nonumber
\\[0.2cm]
\hat{B}_{2} + i \hat{C}_{2} = \sqrt{2} \; (L_{2} -M_{3}) \; ,
\nonumber
\\[0.2cm]
\hat{B}_{2} - i \hat{C}_{2} = \sqrt{2} \; (\bar{M}^{*}_{3} - N^{*}_{1}) \; ,
\nonumber
\\[0.2cm]
\hat{B}_{3} + i \hat{C}_{3} = \sqrt{2}  \; (\bar{M}_{3} - N_{1}) \; ,
\nonumber
\\[0.2cm]
\hat{B}_{3} - i \hat{C}_{3} = \sqrt{2}  \;(L^{*}_{2} - M^{*}_{3})  \;,
\label{43}
\end{eqnarray}
\noindent from here it follows
\begin{eqnarray}
\hat{B}_{0} = 2^{-1/2} ( L_{3} - M_{1} + L^{*}_{3} - M^{*}_{1}) \; ,
\nonumber
\\[0.2cm]
\hat{B}_{1} = 2^{-1/2} ( \bar{M}_{2} - N_{3} +
\bar{M}^{*}_{2} - N^{*}_{3}) \; ,
\nonumber
\\[0.2cm]
\hat{B}_{2} = 2^{-1/2} ( L_{2} - M_{3} + \bar{M}^{*}_{3} - N^{*}_{1}) \; ,
\nonumber
\\[0.2cm]
\hat{B}_{3} = 2^{-1/2} ( \bar{M}_{3} - N_{1} +
L^{*}_{2} - M^{*}_{3}) \; ,
\nonumber
\\[0.2cm]
i \hat{C}_{0} = 2^{-1/2} ( L_{3} - M_{1} - L^{*}_{3} + M^{*}_{1}) \; ,
\nonumber
\\[0.2cm]
i\hat{C}_{1} = 2^{-1/2} ( \bar{M}_{2} - N_{3} -
\bar{M}^{*}_{2} + N^{*}_{3}) \; ,
\nonumber
\\[0.2cm]
i \hat{C}_{2} = 2^{-1/2} ( L_{2} - M_{3} - \bar{M}^{*}_{3} + N^{*}_{1}) \; ,
\nonumber
\\[0.2cm]
i\hat{C}_{3} = 2^{-1/2} ( \bar{M}_{3} - N_{1} -
L^{*}_{2} + M^{*}_{3}) \; .
\label{43}
\end{eqnarray}

\noindent As it should be expected, the  $\hat{B}_{0},\hat{B}_{1}$ and
$\hat{C}_{0},\hat{C}_{1}$  are real-valued, but $\hat{B}_{3} = \hat{B}^{*}_{2} ,
\;    \hat{C}_{3} = \hat{C}^{*}_{2}$.

\section{Newman-Penrose coefficients in spinor approach }

To present time, the  method of spin coefficients proposed by Newman and Penrose
has  become prevalent in studying  gravitational fields and
particle fields on curved space-time  background (see, for example,
the textbook by Rindler and
Penrose [4] or  surveys by Frolov  [6],   Alekseev and Khlebnikov [7]).
The  point we are going to elaborate below is that the spin coefficients provide us with
gauge non-invariant characteristics of a gravitational field. So the  question of their gauge
properties is of special  physical meaning,  important conceptually and
relevant to ordinary day-to-day technical work with gravitational fields.

Let us  begin with special definition of spin coefficients in spinor approach.
Starting from ordinary (non-isotropic) tetrad and conventional Ricci rotation
coefficients
\begin{eqnarray}
e ^{\beta }_{(x)}(x) \;  , \qquad
\gamma _{abc}(x) = - \; e ^{\beta }_{(a);\nu ;\mu }\; e^{\nu }_{(b)}\;
e^{\mu }_{(c)} \; ;
\nonumber
\end{eqnarray}

\noindent
instead of  $\gamma _{abc}(x)$  one can introduce two complex-valued
 (say "spinor")  objects
$\gamma (x)$  and  $\bar{\gamma }(x)$:
\begin{eqnarray}
\gamma (x) = \left ( \; \bar{\sigma }^{a} \; \sigma ^{b} \otimes
\sigma ^{c} \right ) \; \gamma _{abc}(x) \;  ,\qquad
\bar{\gamma }(x) = \left (\; \sigma ^{a} \; \bar{\sigma }^{b} \otimes
\bar{\sigma }^{c} \right ) \; \gamma _{abc}(x) \; ;
\label{b2.1a}
\end{eqnarray}

\noindent $\gamma (x)$ and $\bar{\gamma }(x)$  stand for  objects with four 2-spinor indices.
Inversely, initial Ricci coefficients can be reconstructed as follows
\begin{eqnarray}
\gamma _{klm}(x) = A \;
 [  \; \mbox{Tr} ( \bar{\sigma }_{k} \; \sigma _{l} \otimes
\mbox{Tr} ( \bar{\sigma }_{n}
\;  ] \; \gamma (x) \;
+  \; A \;
[ \;  \mbox{Tr}  ( \sigma _{k} \bar{\sigma }_{l} \otimes
\mbox{Tr}  ( \sigma _{n} \; ]
 \; \bar{\gamma }(x)  \;  .
\label{b2.1b}
\end{eqnarray}

\noindent It readily can be found $A = 1/16$ by substituting
expressions (\ref{b2.1a}) for  $\gamma (x)$  and  $\bar{\gamma }(x)$  into  (\ref{b2.1b}):
\begin{eqnarray}
\gamma _{klm}(x) = A \;
\mbox{Tr} \;( \bar{\sigma }_{k} \; \sigma _{l}  \;  \bar{\sigma }_{a} \;
\sigma _{b}) \; \;
\mbox{Tr} \; ( \bar{\sigma}_{n} \; \sigma _{c} )   \;
\gamma ^{abc}(x) \;
\nonumber
\\[0.2cm]
+ \;
A \; \mbox{Tr}  \; (\sigma _{k} \; \bar{\sigma }_{l}\; \sigma _{a}\; \bar{\sigma }_{b})\;
\mbox{Tr} \; (\sigma _{n} \; \bar{\sigma }_{c}) \; \gamma ^{abc}(x) \;
\nonumber
\end{eqnarray}

\noindent and taking into account the  known formulas for
Pauli matrix traces
\begin{eqnarray}
\mbox{Tr} \;(\bar{\sigma }_{k} \sigma _{l}\bar{\sigma }_{a}
\sigma _{b})
  = 2  ( g_{kl} g_{ab}   -
g_{ka}  g_{lb}  +
g_{kb} g_{la}  - i  \epsilon _{klab} )  ,
\nonumber
\\[0.4cm]
\mbox{Tr} \; (\sigma _{k}\bar{\sigma }_{l} \sigma _{a}
\bar{\sigma }_{b})
= 2 (  g_{kl}  g_{ab} -  g_{ka}  g_{lb}
 + g_{kb} g_{la}   + i \epsilon _{klab}  )  ,
\nonumber
\\
\mbox{Tr}  (\bar{\sigma }_{n}  \sigma _{c} ) =  2 g_{nc} \; , \qquad
 \mbox{Tr}  ( \sigma _{n}  \bar{\sigma }_{c}) = 2 g_{nc} \;  .
\nonumber
\end{eqnarray}

\noindent
Further, with the use of Ricci coefficient definition, spinors
$\gamma (x)$  and  $\bar{\gamma }(x)$ will read
\begin{eqnarray}
\gamma (x) = - \bar{\sigma }_{\alpha ; \beta }(x) \; \sigma^{\alpha}(x) \
\otimes \sigma^{\beta}(x)  \; ,
\qquad
\bar{\gamma }(x) = - \sigma _{\alpha ;\beta }(x)\;
\bar{\sigma }^{\alpha }(x) \otimes  \bar{\sigma }^{\beta }(x) \; .
\label{b2.3}
\end{eqnarray}

\noindent For the following it will be  more convenient instead of  $\gamma (x), \; \bar{\gamma}(x)$
to use slightly modified  spinors  $\Gamma (x)$ and $\bar{\Gamma }(x)$:
\begin{eqnarray}
\Gamma (x) = ( \epsilon  \otimes I  \otimes I  \otimes I )\;
 \gamma (x) \;   ,
 \qquad
\bar{\Gamma }(x) = ( I \otimes \epsilon  \otimes \epsilon  \otimes
\epsilon  ) \; \bar{\gamma }(x) \; ,\;
\nonumber
\end{eqnarray}

\noindent where  $\epsilon  = - \; i\; \sigma ^{2}$.
These spinors are symmetrical over two first indices:
\begin{eqnarray}
\Gamma (x) =   [\; \gamma _{(\alpha \beta) \dot{\rho} \sigma}(x) \;
 ] \; ,\;
\bar{\Gamma }(x) =  [
\; \bar{\gamma }_{(\dot{\alpha}\dot{\beta})\rho\dot{\sigma}}(x)
\;]\; .
\nonumber
\end{eqnarray}

\noindent
 Explicit spinor-indices structure of all other quantities involved  looks  as
follows
\begin{eqnarray}
\sigma ^{a} = \sigma ^{a} _{\dot{\alpha} \beta}  \; , \qquad
\bar{\sigma }^{a}  = \bar{\sigma }^{a \alpha \dot{\beta}} \; , \qquad \qquad
\nonumber
\\[0.2cm]
\gamma (x) =  ( \gamma ^{\alpha }_{\beta \dot{\rho} \mu} (x)  ) =
[ \;(\; \bar{\sigma }^{(a)\dot{\alpha} \nu } \; \sigma ^{b} _{\dot{\nu} \beta} \; ) \;
\sigma ^{c}_{\dot{\rho} \mu} \; \gamma _{abc}(x) \;  ] \;,
\nonumber
\\[0.2cm]
\bar{\gamma }(x) =  ( \gamma _{\dot{\alpha}} ^{\dot{\;\;\beta}
\rho \dot{\mu}}(x)  ) =  [ \; ( \sigma ^{a}_{\dot{\alpha} \nu}\;
\sigma ^{b \;\nu \dot{\beta}}  )  (\sigma )^{c\;\rho \dot{\mu}}
\gamma _{abc}(x) \;  ] \; ,
\nonumber
\\[0.2cm]
\sigma ^{\nu }(x) = (\; e ^{nu}_{\dot{\alpha} \beta}(x) \; ) =
 [ \; \sigma ^{a}_{\dot{\alpha} \beta}\; e^{nu}_{(a)} (x) \; ] \; , \qquad
\nonumber
\\[0.2cm]
\bar{\sigma }^{\nu }(x) = ( \; e^{\nu \dot{\alpha} \beta} (x) \; ) =
[\; \bar{\sigma }^{a\; \alpha \dot{\beta}}\; e ^{\nu}_{(a)} (x) \;] \; . \;\;\;\;
\nonumber
\end{eqnarray}

 Now, with the  use of special letter-notation
(\ref{3.5})  for $\Gamma (x)$ and   $\bar{\Gamma }(x)$ one gets
\begin{eqnarray}
\Gamma  = -   \left ( \begin{array}{cc}
2l_{\alpha ;\beta }  \bar{m}^{\alpha } &
l_{\alpha ;\beta }   n^{\alpha } +
m_{\alpha ;\beta } \bar{m}^{\alpha }    \\[0.2cm]
  l_{\alpha ;\beta }  n^{\alpha }  +
m_{\alpha ;\beta } \bar{m} ^{\alpha }  &
 - 2n_{\alpha ;\beta }  \bar{m} ^{\alpha }
\end{array} \right ) \otimes
\left ( \begin{array}{cc}
l^{\beta }  &   m^{\beta }  \\[0.2cm]
 \bar{m}^{\beta }  &  n^{\beta }
\end{array} \right )  \; ,
\nonumber
\\[0.6cm]
\bar{\Gamma } = -  \left ( \begin{array}{cc}
2l_{\alpha ;\beta }  m^{\alpha }   &
 l_{\alpha ;\beta } n^{\alpha }
+ \bar{m} _{\alpha ;\beta } m^{\alpha }  \\[0.2cm]
 l_{\alpha ;\beta }  n^{\alpha }  +
\bar{m}_{\alpha ;\beta }  m ^{\alpha }  &
-2n_{\alpha ;\beta }  m^{\alpha }
\end{array} \right )
 \otimes
\left ( \begin{array}{cc}
l^{\beta }   &   \bar{m}^{\beta } \\[0.2cm]
m^{\beta }   &   n^{\beta }       \end{array} \right )  \; .
\nonumber
\end{eqnarray}

It should be  noted $(\Gamma (x))^{*}  = \bar{\Gamma }(x)$, where the symbol $*$
stands for  complex  conjugation. Therefore one may consider transformation law for
the  $\Gamma (x)$ only. Besides,
you may omit indices $(\alpha ,\beta )$  for short -- then you
arrive at
\begin{eqnarray}
\Gamma (x)  = - 2  \left ( \begin{array}{cc}
l  \bar{m}     &     {1 \over 2}  ( l n + m  \bar{m} ) \\[0.2cm]
{1 \over 2}  ( l  n + m  \bar{m} )      &    -n  \bar{m}
\end{array} \right )
\otimes
\left ( \begin{array}{cc}
l & m   \\[0.2cm]
\bar{m}  & n       \end{array} \right )  \; .
\label{b2.6c}
\end{eqnarray}

\noindent All 12 independent components of the $\Gamma (x)$
(the  first matrix in  (\ref{b2.6c}) is  symmetrical one) may be
listed  in  the manner we  like: with the  help of 12  different letter symbols
-- see (\ref{3.10}).
 The  quantities  $( L_{i}, \; N_{i}, \;M_{i}, \;\bar{M}_{i} )$
may be straightforwardly connected with the so-called
Newman-Penrose spin coefficients $( k , \;\pi  , \; \epsilon  ; \;
\rho  , \; \lambda  , \; \alpha  ; \; \sigma  , \; \mu , \; \beta  ; \;
\tau , \; \nu , \; \gamma  )$:
\begin{eqnarray}
\left. \begin{array}{ccc}
L_{1} \;  = \;  c \; k^{*}    \qquad     &
L_{2} \;  = \;  c \; \pi ^{*} \qquad     &
L_{3} \;  = \;  c \; \epsilon ^{*}      \\[0.2cm]
N_{1} \;  = \;  c \; \tau^{*} \qquad     &
N_{2} \;  = \;  c \; \nu ^{*} \qquad     &
N_{3} \;  = \;  c \; \gamma ^{*}        \\[0.2cm]
M_{1} \;  = \;  c \; \rho ^{*} \qquad    &
M_{2} \;  = \;  c \; \lambda ^{*} \qquad &
M_{3} \;  = \;  c \; \alpha ^{*}        \\[0.2cm]
\bar{M}_{1} \; = \;  c \; \sigma ^{*} \qquad  &
\bar{M}_{2} \; = \;  c \; \mu ^{*} \qquad     &
\bar{M}_{3} \; = \;  c \; \beta ^{*}
\end{array} \right.
\nonumber
\end{eqnarray}

\noindent where  $c = 2^{3/2}$.

\section{ Gauge transformation}

\hspace{5mm}
Now the task is  to consider the  problem
of general 6-parametric  gauge  transformations for spin coefficients
 $( L_{i} , \; N_{i} , \; M_{i} , \; \bar{M}_{i} )$
under local Lorentz group. Let us start with the  known
gauge law for the  ordinary Ricci object $\gamma _{abc}(x)$:
\begin{eqnarray}
\gamma '_{abc}(x) =
 L^{\;\;k}_{a}(x) \;  L^{\;\;l}_{b}(x) \;  L^{\;\;n}_{c}(x) \;
\gamma _{kln}(x)  \;
\nonumber
\\[2mm]
+ \; L^{\;\; k}_{a}(x)  \; g_{kl} \; \left [ {\partial \over \partial x^{\mu}}
 L^{\;\;l}_{b}(x)  \right ] \;   L^{\;\;n}_{c}(x) \;
e ^{\mu }_{(n)}(x)  ] \; .
\nonumber
\end{eqnarray}

\noindent
As  the $4\times 4 $  Lorentz matrix   $L^{\;\;b}_{a}(x)$ depends upon
both $k(x)$  and conjugate  $k^{*}(x)$,
the second term in  the formula  contains both $(\partial /\partial x^{\mu })  \; k_{a}$  and
$(\partial /\partial x^{\mu })  \; k^{*}_{a}$.

It will be  seen, simplification we  arrive at in spinor  basis $\gamma (x)$ and
 $\bar{\gamma }(x)$ is that each of spinors $\gamma (x)$ and
 $\bar{\gamma }(x)$  transforms independently within  itself,
besides terms $(\partial /\partial x^{\mu })  \; k_{a}$  and
$(\partial /\partial x^{\mu })  \; k^{*}_{a}$ enter its own transformation law.
The task is to  find  these two formulas.

Let us proceed from spinors
$\gamma' (x)$  and     $\bar{\gamma}'(x)$ in a primed tetrad
$e^{'\alpha} _{(a)}(x)  = L^{\;\;b}_{a}(x) \;
e^{\alpha}_{(b)}(x)$:
\begin{eqnarray}
\gamma '(x) = - \bar{\sigma }' _{\alpha ; \beta }\;
        \sigma ^{'\alpha } \otimes \sigma ^{'\beta } \;  ,
\qquad
\bar{\gamma }'(x) = - \sigma '_{\alpha ; \beta }\;
\bar{\sigma }^{'\alpha} \otimes   \bar{\sigma }^{'\beta} \; .
\label{3.2 }
\end{eqnarray}

\noindent Taking the   the known relations [5]
\begin{eqnarray}
\bar{\sigma }^{'\alpha}(x)  = B(k(x)) \; \bar{\sigma }(x) \; B(k^{*}(x)) \; ,
\qquad
\sigma ^{'\alpha}(x)  = B(\bar{k}^{*}(x))\; \sigma (x)\; B(\bar{k}(x)) \; , \;
\nonumber
\end{eqnarray}

\noindent we  get to
\begin{eqnarray}
\gamma '(x)
= -  \left  [\;
{\partial B(k) \over \partial x^{\mu }}
(\bar{\sigma }^{\nu }  \sigma _{\nu })  B(\bar{k})  +
B(k)  \bar{\sigma }_{\nu ;\mu }  \sigma ^{\nu } B(\bar{k})  \right.
\nonumber
\\
\left.  +  B(k) \bar{\sigma}_{\nu} {\partial B(k^{*}) \over \partial x^{\mu }}
B(\bar{k}^{*}) \sigma^{\nu} B(\bar{k})   \right ] \otimes
 B(\bar{k}^{*}) \sigma ^{\mu}  B(\bar{k})    ,
\nonumber
\\[0.2cm]
\bar{\gamma }'(x)
  = - \left  [
{\partial B(\bar{k}^{*}) \over \partial x^{\mu }}
( \sigma ^{\nu }  \bar{\sigma }_{\nu } ) B(k^{*})  +
B(\bar{k}^{*}) \sigma _{\nu ;\mu }  \bar{\sigma }^{\nu } B(k^{*})
\right.
\nonumber
\\
\left.
 +  B(\bar{k}^{*})  \sigma _{\nu }
{\partial  B(\bar{k})  \over \partial x^{\mu }}
B(k )  \bar{\sigma }^{\nu} B(k^{*}) \right ]
\otimes   B(k ) \bar{\sigma }^{\mu }  B(k^{*})   .
\nonumber
\end{eqnarray}

\noindent In both formulas third terms vanish because of
two identities
\begin{eqnarray}
\bar{\sigma }_{\nu }\;  [\;
{\partial B(k^{*}) \over \partial x^{\mu }} \;  B(\bar{k}^{*})\; ]\;
 \sigma ^{\nu } \equiv  0 \; ,
 \qquad
 \sigma _{\nu } \; [\; {\partial  B(\bar{k}) \over \partial x^{\mu }}  \;
B(k) \;] \;  \bar{\sigma }^{\nu } \equiv  0 \;
\nonumber
\end{eqnarray}

\noindent hold. So that we arrive at
\begin{eqnarray}
\gamma '(x)  =   (
 B(k) \otimes  \tilde{B}(\bar{k}) \otimes  B(\bar{k}^{*})  \otimes
\tilde{B}(\bar{k})
 )  \gamma (x)
+ 4 \left (
B(k) {\partial  B(\bar{k})  \over \partial x^{\mu}} \right )
\otimes  \;  B(\bar{k}^{*})  \sigma ^{\mu }  B(\bar{k})   \; ,\qquad
\nonumber
\\[0.2cm]
\bar{\gamma }'(x) =  [ (
B(\bar{k}^{*}) \otimes \tilde{B}(k^{*}) \otimes B(k) \otimes \tilde{B}(k^{*})
)   \bar{\gamma }(x)
+ 4 \;  \left (B(\bar{k}^{*}) {\partial  B(k^{*}) \over \partial x^{\mu}}  \right )
\otimes  \;  B(k)  \bar{\sigma }^{\mu } B(k^{*})   ]  .
\nonumber
\end{eqnarray}

\noindent From this, with the use of
$$
\epsilon \; B(k)  = \tilde{B} (\bar{k}) \; \epsilon \;  , \;\;
(\; \epsilon  _{\dot{\mu} \dot{\nu}} =  - i\; \sigma ^{2} \; ;
\; \epsilon _{\mu\nu} = - i \; \sigma ^{2} ) \;  ,
$$

\noindent gauge formulas for $\Gamma (x)$  and $\bar{\Gamma }(x)$ follow
\begin{eqnarray}
\Gamma ' (x)
=   (
 \tilde{B}(\bar{k}) \otimes  \tilde{B}(\bar{k})
\otimes B(\bar{k}^{*}) \otimes \tilde{B}(\bar{k})
)   \Gamma (x)
+  4 \epsilon    (
B(k) {\partial B(\bar{k}) \over \partial x^{\mu}}     )
\otimes   (
B(\bar{k}^{*})  \sigma ^{\mu } B(\bar{k}  ) )  , \qquad
\label{b3.4a}
\end{eqnarray}
\begin{eqnarray}
\bar{\Gamma }'(x)
=   (
 B(\bar{k}^{*}) \otimes B(\bar{k}^{*}) \otimes
\tilde{B}(\bar{k}) \otimes B(\bar{k}^{*})   )  \bar{\Gamma }(x)
+  4  ( B(\bar{k}^{*}) {\partial  B(k^{*})  \over \partial x^{\mu} }
 \epsilon  ) \otimes  (\epsilon  B(k)\bar{\sigma }^{\mu }
B(k^{*})   ) \epsilon      .
\label{b3.4b}
\end{eqnarray}

\noindent In more  detailed form  the law  (\ref{b3.4a})  means
\begin{equation}
\begin{split}
L'_{i} =  b b^{*} F_{i}
                 -  d d^{*}  G_{i}
                 - d  b^{*} H_{i}
                 -  d^{*} b \Delta _{i}
\\[0.2cm]
M'_{i}  =   -  c  b^{*}  F_{i}
                 - a d^{*}  G_{i}
                 +  a b^{*}  H_{i}
                 +   c  d^{*}  \Delta _{i}
\end{split}
\label{b3.5a}
\end{equation}
\begin{equation}
\begin{split}
\bar{M}' _{i}   =    -  c^{*} b   F_{i}
                          - a^{*}  d   G_{i}
                          +  d  c^{*}   H_{i}
                          +  a^{*}  b   \Delta _{i}
\\[0.2cm]
{N}' _{i}   =   c  c^{*}   F_{i}
                          + a  a^{*}   G_{i}
                          - a  d^{*}   H_{i}
                          -  a^{*}  c  \Delta _{i}
\end{split}
\label{b3.5b}
\end{equation}

\noindent  where
\begin{eqnarray}
F_{1} = [ ( b^{2} L_{1} + d^{2} L_{2} - 2 b d L_{3} )
- \; 2 l^{\mu } ( b \partial _{\mu } d - d \partial _{\mu } b ) ] \; ,
\nonumber
\\
G_{1} = [ ( b^{2} N_{1} + d^{2} N_{2}-  2 b d N_{3} )
- \;2 n^{\mu }
( b \partial _{\mu } d -  d \partial _{\mu } b ) ] \; ,
\nonumber
\end{eqnarray}
\begin{eqnarray}
H_{1} = [ ( b^{2}  M_{1} + d^{2} M_{2} - 2 b d M_{3} )
-\; 2 m^{\mu } ( b\partial _{\mu }d - d \partial _{\mu } b ) ] \; ,
\nonumber
\end{eqnarray}
\begin{eqnarray}
\Delta _{1} = [ ( b^{2} \bar{M}_{1} + d^{2} \bar{M}_{2}
- \;
2 b d \bar{M}_{3} ) -  2\bar{m}^{\mu } (b\partial _{\mu } d  -  d
\partial _{\mu } b )  ] \; ;
\nonumber
\\
F_{2} = [ ( c^{2} L_{1}  + a^{2}  L_{2}
- \; 2  a c L_{3} ) - 2 l^{\mu } (c\partial _{\mu } a - a
\partial _{\mu } c ) ] \;,
 \nonumber
\\
G_{2} = [  ( c^{2}N_{1}+a^{2}N_{2} - 2a c N_{3})
 - \; 2 n^{\mu }(c\partial _{\mu } a - a \partial _{\mu } c )] \; ,
\nonumber
\\
H_{2}  = [ (c^{2} M_{1} + a^{2} M_{2}- 2a c M_{3} )
- \; 2 m^{\mu } (c\partial _{\mu }  a - a \partial _{\mu } c ) ] \; ,
\nonumber
\\
\Delta _{2} = [(c^{2}\bar{M}_{1}+ a^{2} \bar{M}_{2} -
2ac \bar{M}_{3} )
 - \; 2\bar{m}^{\mu }(c \partial _{\mu } a - a \partial _{\mu } c )  ]\; ,
\nonumber
\\
F_{3}= [ - b  c L_{1} - a d L_{2} +  ( a b + c d )
 L_{3}
- \; 2 l^{\mu } ( a  \partial _{\mu } b  -  c
 \partial _{\mu } d ) ]\;  ,
 \nonumber
\\
G_{3} = [ - b c N_{1}  - a d N_{2} + ( a  b  + c d ) N_{3}
- \; 2 n^{\mu }(a\partial _{\mu } b  - c \partial _{\mu } d )] \;  ,
\nonumber
\\
H_{3} = [- b c M_{1}  - a  d  M_{2} +  ( a  b  + c d ) M_{3}
-\; 2 m^{\mu } (a \partial _{\mu } b - c \partial _{\mu } d) ] \; ,
\nonumber
\\
\Delta _{3} = [ - b c \bar{M}_{1} - a d \bar{M}_{2}
+(a b + c d )\bar{M}_{3}
 - \;  2 \bar{m}^{\mu } (a \partial _{\mu } b  -  c
\partial _{\mu } d  ) ] \; ;
\nonumber
\end{eqnarray}

\noindent
The  quantities  $( a, \; b,\; c,\; d )$ are elements  of spinor
$(2 \times 2)$ -- transformation $B(k(x)) \in SL(2.C)$ corresponding to the
local Lorentz  matrix:
\begin{eqnarray}
e^{'\mu }_{(b)}(x)  =  L^{\;\; a}_{b}(k(x),\; k^{*}(x))\; ,
\qquad
B(k(x)) =
\begin{pmatrix}{}
 a(x) & c(x) \\
d(x) & b(x)
\end{pmatrix}
.
\label{b3.5c}
\end{eqnarray}

\section{Example, spin coefficients of spherical tetrad}

Here one illustrative  example will be considered: let us  find
explicit spin coefficients  in spherical tetrad of Minkowski space
by means  of a  direct gauge  transformation of zero spin coefficients in Cartesian tetrad.
When starting spin coefficient vanish identically, in the above  formulas
(\ref{b3.5a}) and  (\ref{b3.5b}) are present only second non-uniform terms.

For elements $(a,\; b,\; c,\; d)$ of  spinor matrix corresponding to transition from
Cartesian tetrad  to spherical one, we have
\begin{eqnarray}
a  = \cos {\theta \over 2}\;  e^{i\phi /2}   \; , \;
c  = \sin {\theta \over 2} \; e^{-i\phi /2}  \; ,
\;
d = - \sin {\theta \over 2}\; e^{i\phi  /2} \; , \;
b =   \cos {\theta \over 2}\; e^{-i\phi /2} \; .
\label{b4.1}
\end{eqnarray}

\noindent These parameters are functions of variables $(\theta , \; \phi)$,
therefore in (\ref{b3.5a}) and  (\ref{b3.5b})  it is  convenient
index  $\mu $  to refer to  $x^{\mu } = (t,\; r, \; \theta ,\; \phi )$.
We have
\begin{eqnarray}
\sigma ^{\mu }(x)  = ( \partial x^{\mu } / \partial x^{i} )
\sigma ^{i}(x)\; , \;
\sigma ^{i}(x)  = \sigma ^{a} \; e^{i}_{(a)}
\label{b4.2a}
\end{eqnarray}

\noindent where  $x^{\mu }$
 and $x^{i}$ stand for spherical and Cartesian coordinates respectively,
$e^{i}_{(a)} = \delta ^{i}_{a}$ is a Cartesian tetrad
defined in Cartesian coordinates. Correspondingly,  $\sigma ^{\mu }(x)$ is
\begin{eqnarray}
\sigma ^{\mu }(x) =   \left ( \begin{array}{cc}
l^{\mu }(x)  &  m^{\mu}(x) \\[0.2cm]
\bar{m}^{\mu } (x) &  n ^{\mu} (x)   \end{array}  \right )
=  { \partial x^{\mu } \over \partial x^{a}}\;  \sigma ^{a}  .
\label{b4.2b}
\end{eqnarray}

\noindent From (\ref{b4.2b}), with the  relations
\begin{eqnarray}
{\partial r \over \partial x}  = \sin \theta  \cos \phi   ,
{\partial r \over \partial y} = \sin \theta  \sin \phi   ,
{\partial r \over   \partial z} = \cos \theta  ,
\nonumber
\\[0.2cm]
{\partial \theta \over  \partial x} = { \cos \theta \cos \phi \over r}  ,
{\partial \theta \over  \partial y} = { \cos \theta  \sin \phi \over r} ,
{\partial \theta \over  \partial z} = {- \sin \theta \over r}  ,
\nonumber
\\[0.2cm]
{\partial \phi \over \partial x} =  {- \sin \phi  \sin \theta \over r}   ,
{\partial \phi \over \partial y} = { \cos \phi  \sin \theta \over r} ,
{\partial \phi \over \partial z} = 0
\nonumber
\end{eqnarray}

\noindent it follows  $(l^{\mu } , \; n^{\mu } , \;
m^{\mu },\; \bar{m}^{\mu } )$:
\begin{eqnarray}
l^{\mu } =  \left ( \begin{array}{c}
1  \\[0.2cm]
  \cos  \theta  \\[0.2cm]
 - {1 \over r} \;\sin  \theta  \\[0.2cm]
 0
 \end{array} \right )
,
m^{\mu } = \left ( \begin{array}{c}
0 \\[0.2cm]
\sin  \theta \; e^{-i \phi}  \\[0.2cm]
{1 \over r} \; \cos  \theta \; e^{-i\phi }  \\[0.2cm]
-{i \over r} \; e^{-i\phi }\;\sin \theta
\end{array} \right ) \; ,
\nonumber
\\[0.2cm]
n^{\mu } =   \left ( \begin{array}{c}
1 \\[0.2cm]
-\cos  \theta  \\[0.2cm]
+ {1 \over r} \; \sin  \theta   \\[0.2cm]
0
\end{array} \right )
 ,
\bar{m}^{\mu } = \left ( \begin{array}{c}
0 \\ \sin  \theta  e^{+i\phi } \\[0.2cm]
{1 \over r} \; \cos  \theta \; e^{+i\phi }
 \\[0.2cm]
+ {i \over r}\; e^{+i\phi } \; \sin \theta
 \end{array} \right )  .
\nonumber
\end{eqnarray}

\noindent In addition
\begin{eqnarray}
b  {\partial d \over \partial \theta } -  d {\partial b \over \partial \theta}
= - {1 \over 2}  , \qquad
c { \partial a \over \theta }  -  a {\partial c \over \theta }  =
- {1 \over 2}  ,
\nonumber
\\[0.2cm]
b {\partial  d \over \partial \phi }  -  d {\partial b \over \partial \phi}
= -{i \over 2}  \sin \theta  ,\;
c{\partial a \over \partial \phi }  -  a {\partial c \over \partial \phi } = +{i \over 2}
 \sin \theta   ,
\nonumber
\\[0.2cm]
a{\partial b \over \partial \theta}  - b{\partial a \over \partial \theta } = 0   , \qquad
a {\partial  b \over \partial \phi }  -  b{\partial a \over \partial \phi}
= - {i \over 2}  \sin \theta    .
\nonumber
\end{eqnarray}

\noindent With the  use of  which we get to the $F_{i}, \;G_{i}, \;H_{i}, \; \Delta _{i}$:
\begin{eqnarray}
F_{1}  = -  { 1 \over r} \; \sin \theta  \; , \;
F_{2} = -\; {1 \over r} \; \sin \theta \; ,
\qquad
G_{1} = +\; {1 \over r} \; \sin \theta \; ,   \;
G_{2} = +\; {1 \over r} \; \sin \theta \; ,
\nonumber
\\[0.2cm]
H_{1} = {1 \over r} \; (+1\; + \; \cos \theta ) \; e^{-i\phi } \; ,
\qquad
H_{2} = {1 \over r} \; (-1\; + \; \cos \theta ) \; e^{-i\phi } \; ,
\nonumber
\\[0.2cm]
\Delta _{1} = {1 \over r} \; (-1 \; + \; \cos \theta ) \; e^{+i\phi } \; ,
\qquad
\Delta _{2} = {1 \over r} \; (+1 \; + \; \cos \theta ) \; e^{+i\phi }  \; ,
\nonumber
\\[0.2cm]
F_{3}= 0 \; , \qquad G_{3} = 0 \; ,
\qquad
H_{3}= +\; {1 \over r}  \; \cot  \theta \; e^{-i\phi }   \; ,
\qquad
\Delta _{3} = -\; {1 \over r} \; \cot  \theta \;  e^{+i\phi } \; .
\nonumber
\end{eqnarray}

\noindent Now, from  (\ref{b3.5a}) and  (\ref{b3.5b})
we can readily  calculate spherical spin coefficients:
\begin{eqnarray}
L_{i} = 0       \;  , \qquad N_{i} =  0   \;  ,
\nonumber
\\[0.2cm]
M_{1} = {2 \over r}  \; , \; M_{2} = 0  \; ,  \;
M_{3} = {1 \over r} \; \cot  \theta  \; ,
\nonumber
\\[0.2cm]
\bar{M}_{1} = 0 \; , \; \bar{M}_{2} = {2 \over r} \; , \;
\bar{M}_{3}  = {1 \over r}  \;  \cot  \theta  \; .
\nonumber
\end{eqnarray}

These quantities are needed in working with spherical coordinates  and tetrad.
It is seen that  a special gauge  transformation is  responsible for their explicit form.

\section{Conclusions}

\hspace{5mm}
It is shown that only 8 different combinations of the
Ricci coeffici\-ents are involved  in the Dirac equation on a curved space-time background.
In other words, $S =1/2$ particle effectively  observes only
'a third' of geometric characteristics of any space-time model,
and by no means responds to remaining 16 ones.
These eight ones may be collected in two 4-vectors $B_{a}(x)$ and $C_{a}(x)$ under local
Lorentz group which has  status of the gauge symmetry group.
In all orthogonal coordinates  one  of these  vectors, "pseudovector"\hspace{2mm} $C_{a}(x)$,
vanishes identically.  The gauge transformation laws of vectors $B_{a}(x)$ and $C_{a}(x)$
are found explicitly.  The Ricci rotation coefficients,  being exploited in
the generally covariant  linear Dirac equation, assume their very strict and subtle
gauge symmetry properties.
Connection of these   $B_{a}(x)$ and $C_{a}(x)$
with the  known Newman-Penrose coefficients is established.
Insight  into  the Ricci object
in terms of
the two vector fields $B_{\alpha}, C_{\alpha}$ seems deeper and simpler than  the
spin coefficients method.

General study of gauge symmetry in Newman-Penrose formalism is done.
To this end, decomposition of the tensor  $\gamma_{abc}(x)$ into two spinors
 $\gamma(x)$  and $\bar{\gamma}(x)$
 is performed. At this Ricci rotation coefficients are divided into two groups:
12 complex functions  $\gamma(x) = ( \gamma^{\alpha}_{\;\;\beta\dot{\rho}\sigma})$
and 12 conjugated to them $\bar{\gamma}(x) =
( \gamma_{\dot{\alpha}}^{\;\;\dot{\beta}\rho\dot{\sigma}})$.
Components of spinor $\bar{\gamma}(x)$ coincide with
12  spin coefficients by Newman-Penrose
$
\kappa \; , \; \pi \; , \; \epsilon \; , \; \rho \; , \; \lambda \; ,
\; \alpha \; , \;  \sigma \; , \; \beta \; , \; \tau \; , \; \nu \; , \; \gamma \; .
$
For listing  these it is  used a special letter-notation
 $L_{i}\; , \; N_{i} \; , \; M_{i} \; , \; \bar{M}_{i}\; (i = 1,\;2,\; 3 )$.

The  formulas for gauge transformations of spin coefficients under local Lorentz group
are derived on the base of spinor approach. In contrast to a generally accepted treatment
of  gauge  symmetry  in  Newman-Penrose  formalism   with  only  2-parametric Lorentz  matrices
 the formulas obtained  are applicable for any general 6-parametric transformation.
There are given two solutions  to the gauge problem: one in the compact form of transformation
laws for spinors $\gamma(x)$  and $\bar{\gamma}(x)$, and another as detailed description
of the latter in terms of spin coefficients.

\label{last}
\end{document}